\documentclass[3p]{elsarticle}

\begin{document}

\title{Selective Kondo strong coupling \\
in magnetic impurity flat-band lattices}

\author[gust]{Duong-Bo Nguyen}
\author[hnue]{Thanh-Mai Thi Tran}
\author[iop]{Thuy Thi Nguyen}
\author[gust,iop]{Minh-Tien Tran 
}
\address[gust]{Graduate University of Science and Technology, Vietnam Academy of Science and Technology,
Hanoi 100000, Vietnam}
\address[hnue]{Faculty of Physics, Hanoi National Education University,
Hanoi 100000, Vietnam}
\address[iop]{Institute of Physics, Vietnam Academy of Science and Technology,
Hanoi 100000, Vietnam}

\begin{abstract}
The periodic Anderson impurity model on the Lieb lattice is studied by the slave-boson mean-field approximation in the strong interaction limit.
The electron structure of conduction electrons on the Lieb lattice features both the band flatness and soft gap at low energy.
With these features conduction electrons can form both the soft-gap and the molecular Kondo singlets with the magnetic impurities, and
this leads to a competition between
the soft-gap and the molecular Kondo effects in the lattice.
We find a selective Kondo strong coupling, where at selected sites the magnetic impurities are strongly coupled to conduction electrons, and at the remaining sites they are decoupled from the lattice. The selective Kondo strong coupling occurs between the full local moment and the full strong coupling regimes, and  it yields an effective lattice depletion.
At low temperature the selection of the Kondo strong coupling is dominant at those lattice sites, where the local density of states of conduction electrons exhibits the flat-band feature, independently of the impurity parameters. Rich phase diagrams for different model parameters are obtained.
\end{abstract}


\maketitle

\section{Introduction}

Electron correlations in flat-band lattices represent an attractive field that continues to expand its horizon. In flat-band systems the Coulomb interaction between electrons is dominant over the kinetic energy, and as a consequence it can be minimized without any cost in the kinetic energy.
This leads to rich phenomena of electron correlations, for instance the flat-band ferromagnetism or the fractional quantum Hall effect without the Landau levels \cite{Tasakirev,Maksymenko,Sheng,Tang,Sun,Neupert}.
When a magnetic impurity is coupled to electrons of the flat band, the electron correlations at the impurity
become the dominant force that leads the magnetic moment of the impurity to form a molecular Kondo singlet with the spin of a single electron at the flat band \cite{Thuy,Fuldebook,Fulde,RZ}. In the molecular Kondo effect,
all dispersive electrons are not involved in the Kondo singlet formation \cite{Thuy}. This essentially yields an entanglement of two spin qubits in solids due to the band flatness and distinguishes the molecular Kondo effect from the conventional Kondo effect in normal metals \cite{Thuy}. On the other hand, the dispersive electrons in various flat-band lattices often have additional special features at low energy. For instance, on the edge centered square or stacked triangular lattices  the electron structure of the tight-binding model also exhibits the Dirac linear dispersion at low energy  \cite{Shen,Dora}. When a magnetic impurity is coupled to electrons of the special dispersive band, their many-body Kondo problem is also distinguished from the conventional one in normal metals \cite{Bulla,Vojta}. In particular, the soft-gap Kondo problem has a very rich phase diagram \cite{Bulla,Vojta}. In the such way, when a magnetic impurity is embedded in flat-band lattices, the Kondo problem can be originated from different natures, depending on the band properties of conduction electrons, which are coupled to the impurity.

When a flat band lattice is doped with many magnetic impurities, the magnetic moment of these impurities could be coupled with conduction electrons from flat and dispersive bands, and they together can form the Kondo singlets with different natures, for instance, the molecular or the soft-gap Kondo singlets. This leads to a competition between different Kondo singlet formations. One may expect from the competition
at selected sites the magnetic impurities
are strongly coupled to conduction electrons leaving the impurities at other sites are decoupled from the lattice.
This is reminiscent of the orbital-selective Mott insulator, where the narrower orbital band becomes insulating,
while the wider band
is still metallic \cite{Anisimov,Koga,Liebsch,Medici,Tran1,Tran2,Tran3}.
The selective Kondo strong coupling yields an effective lattice depletion.
In strongly correlated electron systems the lattice depletion can lead to intriguing phenomena such as ferrimagnetism
\cite{Lieb,Costa} or the spin liquid behavior \cite{Khatami,Troyer}.
The selective Kondo strong coupling is very similar to the fractionalized Fermi liquid with the breakdown of Kondo screening
\cite{Senthil1,Senthil2}. In the fractionalized Fermi liquid the local moments are partially decoupled from conduction electrons as a result of the breakdown of Kondo screening across the transition from a heavy fermion liquid to a non-magnetic phase. Recently, experiments found a partial Kondo screening of the local moments in heavy fermions on geometrically frustrated lattices \cite{Oyamada,Sakai}.

In this paper we investigate a possibility of the selective Kondo strong coupling in the magnetic impurity flat-band lattices. One of the simplest flat-band lattices is the Lieb lattice, where the electron structure of the tight-binding model features both the band flatness and the Dirac linear dispersion at
low energy \cite{Tasaki,Tien}. With the such electron structure the Lieb lattice allows us to study the competition between the molecular and the soft-gap Kondo strong couplings. As a result of the competition a selective Kondo strong coupling may occur. For the purpose we study the periodic Anderson model (PAM) on the Lieb lattice. The model essentially describes the hybridization between the magnetic impurities and conduction electrons of both the flat and the Dirac-cone bands. The competition between the molecular and the soft-gap Kondo strong couplings could emerge at strong electron correlations. We will adopt the slave-boson mean-field approximation to study the competition \cite{Newns,Coleman}. The slave-boson mean-field approximation is simple, and it can well describe the essential features of the strong coupling(SC) and local moment (LM)
regimes in the PAM \cite{Newns,Coleman}.
Recently, the slave-boson mean-field approximation was also used to study the interlay between the Kondo effect and topology \cite{Dzero,Si}.
We find very rich phase diagrams depending on the impurity parameters. In general, the full local moment (FLM) regime, where all magnetic impurities are decoupled from conduction electrons, appears at high temperature. At low temperature and strong hybridization, the full strong coupling (FSC) regime, where all magnetic moments form the Kondo singlets with conduction electrons, exists. Between the FLM and the FSC regimes, various selective Kondo strong coupling regimes occur. The stability of the selective Kondo strong coupling is essentially due to the low-energy properties of conduction bands, which hybridize with magnetic impurities.
The obtained results predict the selective Kondo strong coupling in heavy-fermion materials, topological semimetals, systems with nonuniform lattice coordination number, where either a flat band exists or the bands of conduction electrons have qualitatively different low-energy properties.

The present paper is organized as follows. In Sec. II we describe the PAM on the Lieb lattice. In this section we also present the slave-boson approach and its mean-field approximation. The numerical results for depleted lattices are presented in Sec. III, and in Sec. IV the phase diagrams for uniform hybridizations are presented. Finally, discussion and conclusion are presented in Sec. V.

\section{Periodic Anderson model on the Lieb lattice and slave-boson mean-field approximation}

\begin{figure}[t]
\centering
\includegraphics[width=0.4\textwidth,bb=113 -33 456 336]{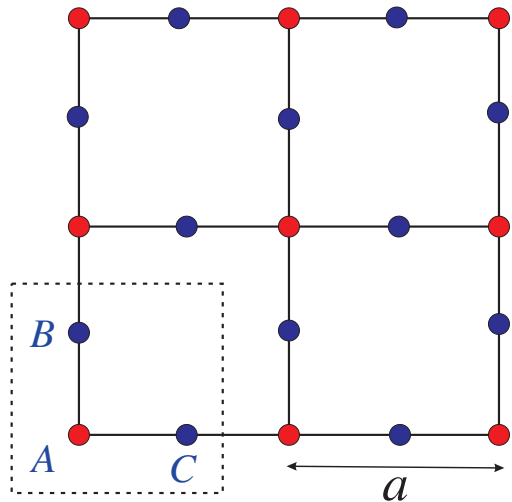}
\vspace{-3cm}
\caption{The Lieb lattice structure.}
\label{fig1}
\end{figure}

The PAM is a lattice generation of the single impurity Anderson model \cite{Anderson}. It describes a lattice of localized electrons hybridized with conduction electrons. The PAM is a suitable model for studying heavy fermion compounds \cite{Fulderev}.
Its Hamiltonian reads
\begin{eqnarray}
H  &  = & -t
\sum\limits_{\left\langle i,j\right\rangle \sigma}
c_{i\sigma}^{\dagger} c_{j\sigma}+
\sum\limits_{i\sigma} \varepsilon_{i}
f_{i\sigma}^{\dagger}f_{i\sigma}+U \sum\limits_{i}
n_{fi\uparrow}^{\dagger}n_{fi\downarrow}\nonumber\\
& &  +
\sum\limits_{i\sigma} V_i
c_{i\sigma}^{\dagger}f_{i\sigma} + \textrm{H. c.} ,
\label{ham}
\end{eqnarray}
where $c^{\dagger}_{i\sigma}$ ($c_{i\sigma}$) is the
creation  (annihilation) operator for conduction electron with spin $\sigma$ at lattice site $i$.
$t$ is the nearest neighbor hopping parameter. $f^{\dagger}_{i\sigma}$ ($f_{i\sigma}$) represents the creation (annihilation) operator for the magnetic impurity with spin $\sigma$ at site $i$. $\varepsilon_{i}$ is the energy level of the magnetic impurity at site $i$. $U$ is  local Coulomb
interaction of impurity electrons. $n_{fi\sigma}=f^{\dagger}_{i\sigma}f_{i\sigma}$ is the impurity number operator.  For simplicity, we assume the magnetic impurities have the spin degeneracy $N_f=2$. $V_i$ is the hybridization strength between conduction electrons and the magnetic impurity at site $i$. In the tight-binding model, the electron structure of conduction electrons depends on the lattice structure. We consider the PAM defined on the Lieb lattice. The Lieb lattice is an edge centered square lattice (see Fig. \ref{fig1}). It is the basic structure of layered cuprates \cite{Tasakirev}. The Lieb lattice is bipartite, but its unit cell contains three lattice sites. The square with three sites $A$, $B$, $C$ can be chosen for the unit cell, as shown in Fig. \ref{fig1}. We take the lattice parameter $a=1$.

Without the magnetic impurities, the tight-binding model on the Lieb lattice features a flat band touching two linearly dispersing bands at the $\mathbf{M}=(\pi,\pi)$ point in the first Brillouin zone. The low energy effective Hamitonian can be obtained by expanding the tight-binding Hamiltonian around the $\mathbf{M}$ point. We obtain at the linear order
\begin{eqnarray}
H(\delta \mathbf{k})=\left(
\begin{array}
[c]{ccc}
0 & -t \; \delta k_{x}   & -t \; \delta k_{y}   \\
-t \; \delta k_{x}   & 0 & 0  \\
-t \; \delta k_{y}   & 0 & 0
\end{array}
\right) = -t \delta\mathbf{k} \cdot \mathbf{S}, \label{spin1}
\end{eqnarray}
where $\delta \mathbf{k}=\mathbf{M}-\mathbf{k}$, and $\mathbf{S}$ is the spin-$1$ matrices
\begin{eqnarray*}
S_x=\left(\begin{array}
[c]{ccc}
0 & 1   & 0  \\
1    & 0 & 0  \\
0   & 0 & 0
\end{array}\right),
S_y=\left(\begin{array}
[c]{ccc}
0 & 0   & 1  \\
0    & 0 & 0  \\
1   & 0 & 0
\end{array} \right),
S_z=\left(\begin{array}
[c]{ccc}
0 & 0   & 0  \\
0    & 0 & -i  \\
0   & i & 0
\end{array}\right),
\end{eqnarray*}
which obey the Lie algebra $[S_\alpha,S_\beta]=i\epsilon_{\alpha\beta\gamma} S_\gamma$.
Equation (\ref{spin1}) shows that the low energy dynamics of the tight-binding model on the Lieb lattice is identical to the one of
the helical spin-$1$.
The effective Hamiltonian has three eigenvalues $E_{\pm}=\pm t |\delta \mathbf{k}|$, $E_0=0$, which correspond to the linearly
dispersing and flat bands touching at the $\mathbf{M}$ point. We will see later, these qualitatively different behaviors of conduction electrons
at low energy are the origin of the selective Kondo strong coupling.
We can refer the $A$, $B$, $C$ sites as the orbital degree of the model. However,
their Bloch functions are not the eigenfunctions of the effective Hamiltonian, and in particular, they are not localized.
The low-energy effective Hamiltonian can be considered as the two-dimensional counterpart of the triply band crossing lattices of topological materials \cite{Bradlyn}. The Lieb lattice can be realized by various ways such as loading ultracold atoms in optical lattices \cite{Taie}, making an array of optical waveguides \cite{Vicencio1,Vicencio2,Mukherjee}, or molecular designing \cite{Slot}. Recently, various proposals for simulating the PAM by loading ultracold atoms in optical lattices have been reported \cite{Kawakami, Zhong}. With the advantages of quantum simulations in optical lattice, one can expect realizations of the PAM on the Lieb lattice too.

When a single magnetic impurity is embedded in the Lieb lattice, the Kondo problem totally depends on the local density of states (LDOS) of conduction electrons that are coupled to the magnetic impurity \cite{Thuy,Bulla}. The LDOS of conduction electrons at sites $B$ or $C$ exhibits both the flat-band and the soft-gap features, whereas at sites $A$ it only features a soft gap \cite{Tasakirev,Thuy}. As a consequence, at sites $B$ or $C$, the magnetic impurity forms the molecular Kondo singlet with a single conduction electron  \cite{Thuy,Fuldebook,Fulde,RZ}, whereas at sites $A$, the soft-gap Kondo effect occurs \cite{Bulla,Vojta}. In the impurity lattice case, when $V_B=V_C=0$, the magnetic impurities hybridize only with conduction electrons at sites $A$, and one may expect that only a lattice soft-gap Kondo effect occurs. In contrast, when $V_A=0$, the magnetic impurities could exhibit a lattice molecular Kondo effect. For finite $V_A$, $V_B$, and $V_C$, there would be a competition between the soft-gap and the molecular Kondo effects in the lattice.

We use the slave boson method to solve the PAM \cite{Newns,Coleman}.
Since the Kondo strong coupling is non-perturbative, we concentrate on the strong correlation region.
In the limit $U \rightarrow \infty$, we represent
\begin{eqnarray*}
f_{i\sigma} \rightarrow \tilde{f}_{i\sigma} b^\dagger_i,
\end{eqnarray*}
where $b^\dagger_i$ is the slave boson, which represents the empty state of the magnetic impurity at site $i$. $\tilde{f}_{i\sigma}$ is the fermion operator for the impurities when their double occupancy is completely excluded. This slave-boson representation enlarges the state space at every site. In order to eliminate the unphysical states, the constrain condition
\begin{eqnarray}
 b^\dagger_i b_i + \sum_\sigma \tilde{f}_{i\sigma}^\dagger \tilde{f}_{i\sigma} =1 \label{cc}
\end{eqnarray}
is imposed. Within the slave-boson representation, Hamiltonian in Eq. (\ref{ham}) becomes
\begin{eqnarray}
H  & = &  -t
\sum\limits_{\left\langle i,j\right\rangle \sigma}
c_{i\sigma}^{\dagger} c_{j\sigma}+
\sum\limits_{i\sigma} \varepsilon_{i}
\tilde{f}_{i\sigma}^{\dagger}\tilde{f}_{i\sigma}  \nonumber\\
&& -  \sum\limits_{i} \lambda_i ( b^\dagger_i b_i + \sum\limits_{\sigma}
\tilde{f}_{i\sigma}^{\dagger}\tilde{f}_{i\sigma})
+
\sum\limits_{i\sigma} V_i
c_{i\sigma}^{\dagger}\tilde{f}_{i\sigma} b^\dagger_i + \textrm{H. c.} ,
\label{sbham}
\end{eqnarray}
where the Lagrange multiplier $\lambda_i$ is introduced for imposing the constrain condition in Eq. (\ref{cc}).
In the mean-field approximation $b_i $ becomes a C-number, and the constrain condition is replaced by its average
\begin{eqnarray}
 |\langle b_i \rangle|^2+ \sum_\sigma \langle \tilde{f}_{i\sigma}^\dagger \tilde{f}_{i\sigma} \rangle =1. \label{constrain}
\end{eqnarray}
The slave-boson Hamiltonian in the mean-field approximation becomes quadratic, and it can analytically be solved.
Minimizing the free energy respected to $\langle b_i \rangle$, we obtain another mean-field equation
\begin{equation}
\lambda_{\alpha}   =
\sum\limits_{\sigma}
\left\vert V_{\alpha}\right\vert ^{2}\int d\omega f(\omega)
\frac{\rho_{\alpha}^{c}(\omega)}{\omega-\varepsilon_{\alpha}+\lambda_{\alpha}} , \label{mf1}
\end{equation}
where $\alpha=A, B, C$, and
$f(\omega)=1/(\exp(\omega/T)+1)$ is the Fermi-Dirac distribution function at temperature $T$. $\rho_{\alpha}^{c}(\omega)$ is the LDOS of conduction electrons at site $\alpha$. The mean-field Eqs.~(\ref{constrain})-(\ref{mf1}) can numerically be solved. When the solution $\langle b_\alpha \rangle \neq 0$ exists, the magnetic moments of impurities at $\alpha$ sites are strongly coupled with conduction electrons. This is the SC regime, where the Kondo singlets between the magnetic impurities and conduction electrons are formed. The solution $\langle b_\alpha \rangle = 0$ indicates the LM regime, where the magnetic impurities at $\alpha$ sites  are decoupled from the lattice.
Depending on the model parameters, the mean field Eqs. (\ref{constrain})-(\ref{mf1}) may have the following solutions:
\begin{enumerate}
\item $\langle b_A \rangle = \langle b_B \rangle = \langle b_C \rangle =0$. All magnetic impurities are effectively decoupled from conduction electrons. This is the FLM state.
\item $\langle b_\alpha \rangle = \langle b_\beta \rangle = 0$ and $\langle b_\gamma \rangle \neq 0$. At $\gamma$ sites the magnetic moments of impurities are strongly coupled with conduction electrons, while at the other sites, they are in the LM regime. We call this state by the $\gamma$-selective strong coupling ($\gamma$-SSC). In this state the impurity $\alpha$ and $\beta$ sites are effectively depleted from the lattice. This depletion is purely a correlation effect. For $\gamma=B$ or $\gamma=C$ the SC
    is characterized by the molecular Kondo singlet formation, because the LDOS of conduction electrons at $B$ and $C$ sites features the band flatness \cite{Thuy}. When $\gamma=A$ the SC is the soft-gap Kondo type, because the LDOS of conduction electrons at $A$ sites exhibits a soft gap \cite{Bulla,Vojta}.
\item $\langle b_\alpha \rangle =0$ and $\langle b_\beta \rangle \neq 0$, $\langle b_\gamma \rangle \neq 0$. At $\alpha$ sites the magnetic impurities  are decoupled from conduction electrons, while at the other sites, they are in the SC regime. We refer to this state as $\beta\gamma$-selective strong coupling ($\beta\gamma$-SSC) state. In this state the impurity $\alpha$ sites  are effectively depleted from the lattice.
\item $\langle b_A \rangle \neq 0$, $\langle b_B \rangle \neq 0$, and $\langle b_C \rangle \neq 0$. All magnetic impurities are in the SC regime. This is the FSC state. In this state both the molecular and the soft-gap Kondo effects occur.
\end{enumerate}

The critical line which is the phase boundary of the boson condensation can be determined from Eq. (\ref{mf1}) by taking the limit
$\langle b_\alpha \rangle \rightarrow 0$. In this limit the magnetic impurities are decoupled from conduction electrons and
$\langle \tilde{f}_{\alpha\sigma}^\dagger \tilde{f}_{\alpha\sigma} \rangle = f(\varepsilon_\alpha-\lambda_\alpha)$. This leads Eq. (\ref{constrain})  to give the solution
$\lambda_\alpha = \varepsilon_\alpha $. Together with Eq. (\ref{mf1}) we obtain
\begin{equation}
\frac{\varepsilon_{\alpha}}{\vert V_{\alpha}\vert ^{2}}    =
\sum\limits_{\sigma}
\int d\omega f(\omega) \frac{\rho_{0\alpha}^{c}(\omega)}{\omega}  , \label{cr}
\end{equation}
where $\rho_{0\alpha}^{c}(\omega)$ is the LDOS of conduction electrons in the lattice with impurity sites $\alpha$ depleted.
Equation (\ref{cr}) determines the critical line of the boson condensation at site $\alpha$.
At zero temperature, $f(\omega)$ vanishes for $\omega >0$. Therefore, Eq. (\ref{cr}) has a solution iff $\varepsilon_\alpha <0$. This indicates the Kondo effect occurs only when the energy level of the magnetic impurities must be below the Fermi level of conduction electrons.
When $\varepsilon_\alpha \geq 0$, the boson condensations cannot occur and magnetic orders instead can be stabilized \cite{Hartman}.
In the Lieb lattice, the LDOS at the corner and edge-center sites feature qualitatively different low-energy properties, and this leads to different values of the critical lines.
If the impurity energy level $\varepsilon_\alpha$ and the hybridization $V_\alpha$ are nonuniform
we would obtain additional differences for the critical lines. These features result in a very rich phase diagram. A similar result would be obtained in any system, where the LDOSs of different orbitals have qualitatively different low-energy behaviors. This can occur in lattices with nonuniform electron coordination numbers such as the body-centered-square lattice, the dice lattice, or in general the decorated, depleted lattices, and quasicrystals.

In numerical calculations  we use the half band width of non-interacting conduction electrons $D=2\sqrt{2} t=1$ as the energy unit. The mean-field equations are solved by the Powell hybrid method \cite{Powell}. We are mainly interested in two cases: i) Depleted lattices, where magnetic impurities are regularly decoupled from conduction electrons. ii) Uniform hybridization, where all magnetic impurities are equally coupled to conduction electrons. All magnetic impurities lie below the Fermi energy level of conduction electrons, i.e. $\varepsilon_\alpha <0$.

\section{Depleted lattice}
\label{secdepl}

\begin{figure}[t]
\centering
\begin{tabular}{cc}
\includegraphics[width=0.45\textwidth]{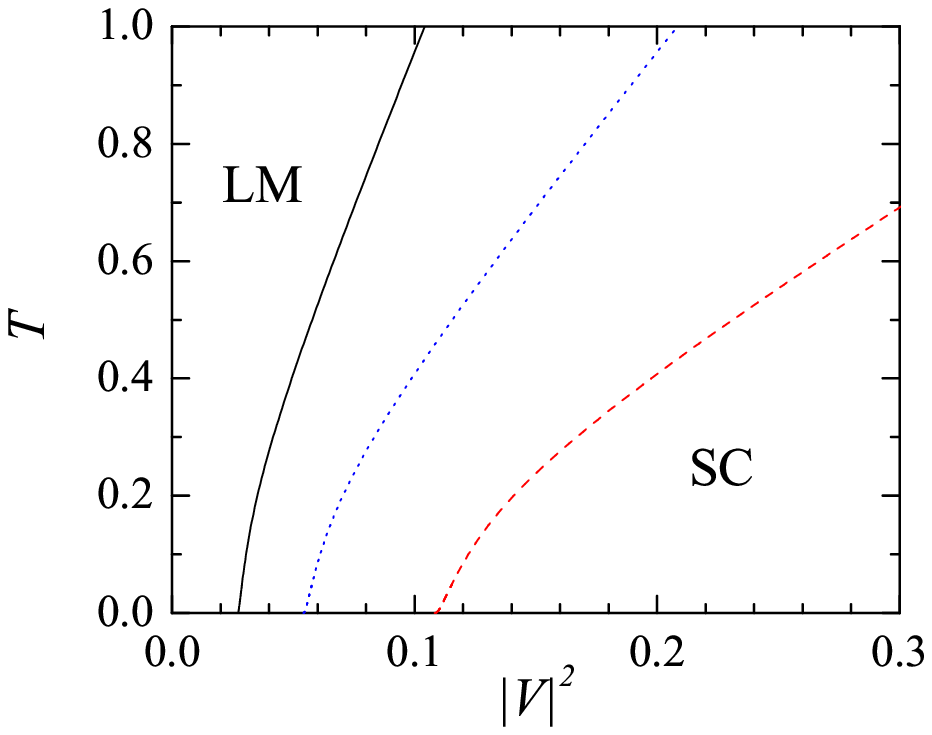}
& \includegraphics[width=0.45\textwidth]{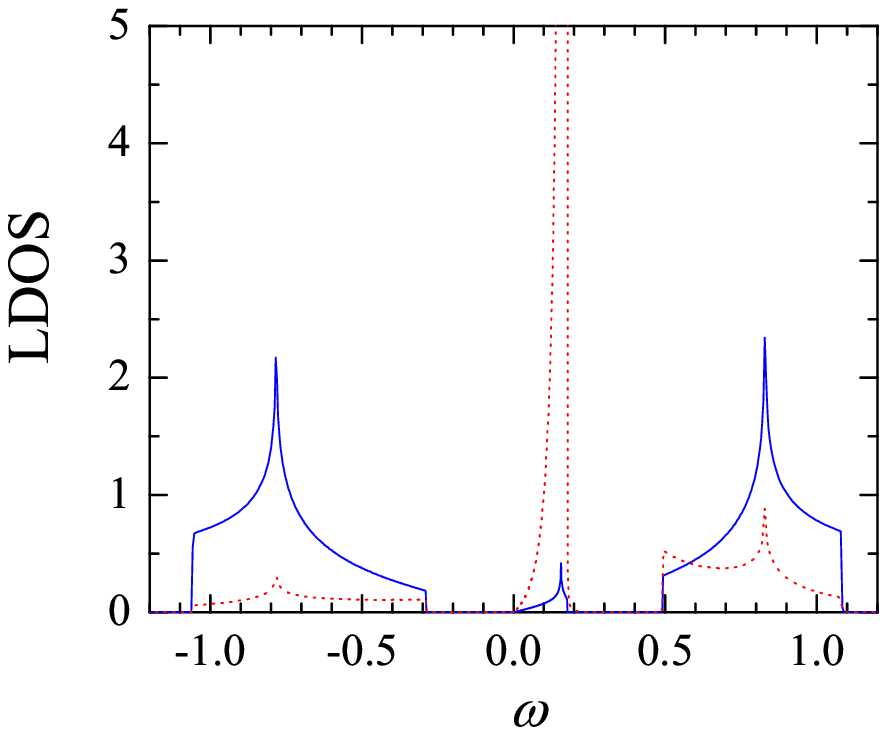}
\end{tabular}
\caption{Depleted lattice with $V_A\equiv V$, and $V_B=V_C=0$. Left panel: Phase diagram for different values of $\varepsilon_A$ (the solid, dotted and dashed lines for $\varepsilon_A=-0.05$, $-0.1$, and $-0.2$, respectively). Right panel: LDOS of conduction electrons (solid line) and of magnetic impurities (dotted line) at $A$ site in the SC phase for $V^2=0.25$, and  $\varepsilon_A=-0.1$ ($T=0.1$). The broadening parameter $\eta=10^{-4}$.}
\label{fig2}
\end{figure}

In this section we consider the depleted lattice, i.e., a regular set of magnetic impurities is decoupled from conduction electrons.
The regular depletion in strongly correlated electron lattices can lead to intriguing phenomena, for instance the ferrimagnetism can occur in the Hubbard model on the $1/4$-depleted square lattice \cite{Lieb}, or in the PAM with half of magnetic impurities regularly depleted \cite{Costa}, the spin liquid behavior occurs in the Hubbard or the Heisenberg models on the $1/5$-depleted square lattice \cite{Khatami,Troyer}.
Here we are mainly interested in two depletion cases: i) $V_A \equiv V \neq 0$, $V_B=V_C=0$; ii) $V_A = 0$, $V_B=V_C\equiv V \neq 0$. In the first case, the sublattice of the magnetic impurities at the edge-center sites is depleted from the whole lattice. Only at the corner sites the hybridization between the magnetic impurities and conduction electrons remains. The LDOS of non-interacting conduction electrons at the corner sites exhibits a vanishing gap $\rho_{A}^{0c}(\omega) \sim |\omega|$, like the one of the Dirac electrons \cite{Vojta}.
This case deals with the soft-gap Kondo problem \cite{Bulla,Vojta}. In Fig.~\ref{fig2} we plot the critical lines which separates the SC phase from the LM one for various values of $\varepsilon_A$. The SC phase exists only when the hybridization $|V_A|$ is larger a critical value $V_{\textrm{cr}}$. The critical value $V_{\textrm{cr}}$ monotonously increases with $|\varepsilon_A|$. In Fig. \ref{fig2} we also plot the LDOS of conduction electrons $\rho_A^{c}(\omega)$ and magnetic impurities $\rho^f_A(\omega)$ at $A$ site   in the SC regime. Since the limit $U \rightarrow \infty$ is the extreme asymmetric case, these LDOSs do not obey the particle-hole symmetry. They exhibit a hybridization band in the gap between two subbands of conduction electrons. The appearance of the hybridization band is due to the condensation of the slave bosons.
The low energy behavior of the LDOS can be derived from the low energy effective Hamiltonian. Expanding the mean-field Hamiltonian around the $\mathbf{M}$ point, we obtain the low energy effective Hamiltonian
\[
H(\delta \mathbf{k})=\left(
\begin{array}
[c]{cccc}
0 & -t \; \delta k_{x}   & -t \; \delta k_{y}   &
V_{A} \langle b_{A}\rangle^{\ast} \\
-t \; \delta k_{x}   & 0 & 0 & 0 \\
-t \; \delta k_{y}   & 0 & 0 & 0 \\
V_{A}^{\ast} \langle b_{A}\rangle & 0 & 0 & \varepsilon_{A}-\lambda_{A}
\end{array}
\right) ,
\]
where $\delta \mathbf{k}=\mathbf{M}-\mathbf{k}$. At low energy $0<\omega/D \ll 1$, we obtain
\begin{eqnarray*}
\rho_A^c(\omega) &=& - \frac{1}{\pi} \sum_{\delta \mathbf{k}} \textrm{Im}  [\omega - H(\delta \mathbf{k})+ i \eta]^{-1}_{11} \sim \omega ,\\
\rho_A^f(\omega) &=& - \frac{1}{\pi} \sum_{\delta \mathbf{k}} \textrm{Im}  [\omega - H(\delta \mathbf{k})+ i \eta]^{-1}_{44} \sim \omega ,
\end{eqnarray*}
where $\eta=0^{+}$ is the broadening small quantity for the Green function.
This low energy behavior of the LDOS is a feature of the soft-gap Kondo problem in the asymmetric SC phase.

\begin{figure}[t]
\centering
\begin{tabular}{cc}
\includegraphics[width=0.45\textwidth]{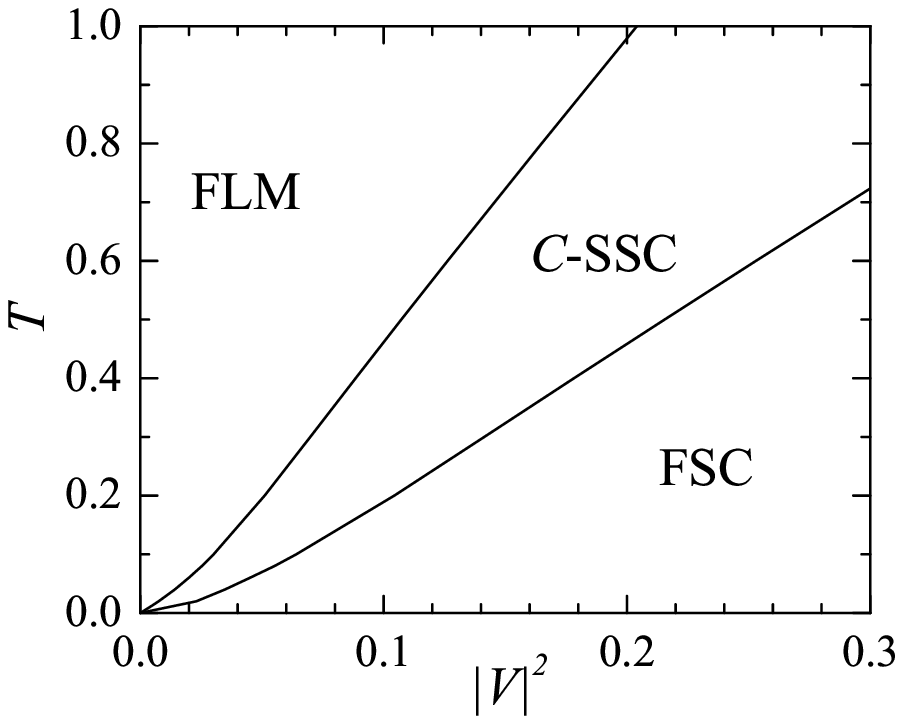} &
\includegraphics[width=0.49\textwidth,bb=0 20 320 276]{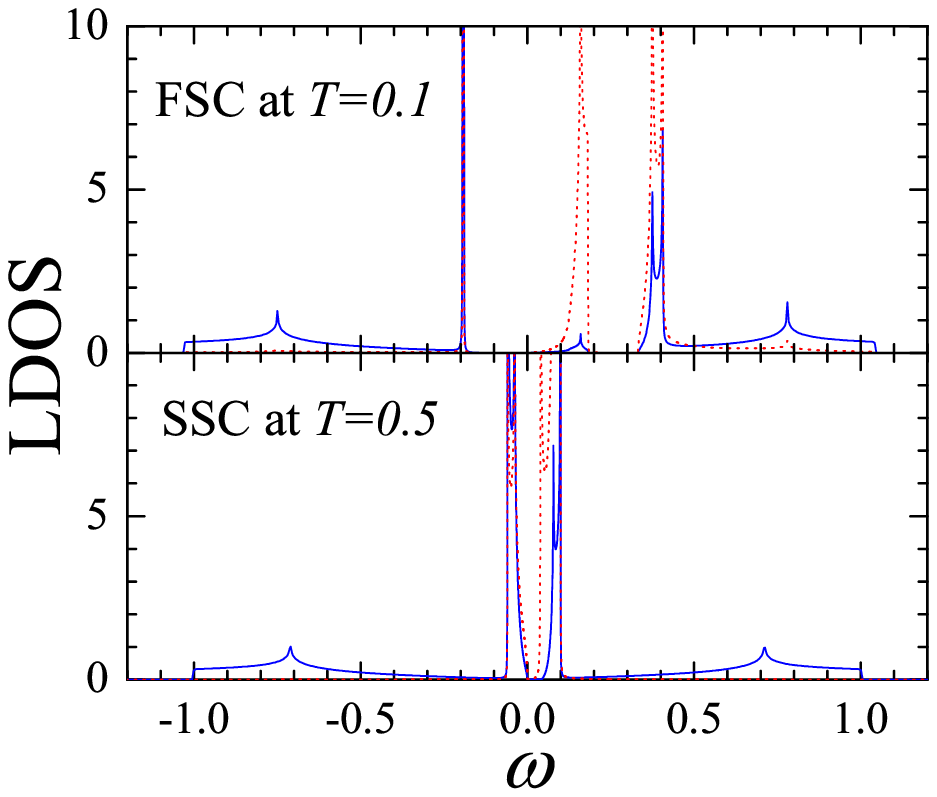}
\end{tabular}
\caption{Depleted lattice with $V_A=0$ and $V_B=V_C\equiv V$. Left panel: Phase diagram for $\varepsilon_B=-0.2$, $\varepsilon_C=-0.1$. Right panel: LDOS of conduction electrons (solid line) and of magnetic impurities (dotted line) at the $C$ site in the FSC and the SSC phases for $V^2=0.15$, $\varepsilon_B=-0.2$, $\varepsilon_C=-0.1$. The broadening parameter $\eta=10^{-4}$.}
\label{fig4}
\end{figure}

In the second depletion case, $V_A = 0$, and $V_B=V_C\equiv V \neq 0$, the sublattice of magnetic impurities at the corner sites is depleted. The LDOS of conduction electrons at the edge-center sites features both the soft gap and the band flatness. This case yields the molecular Kondo problem \cite{Thuy,Fuldebook,Fulde,RZ}. In Fig. \ref{fig4} we plot the phase diagram. When $\varepsilon_B \neq \varepsilon_C$, it displays three different phases. In contrast to the previous depleted case, there is no critical point and the SC phase always exists for any non-zero hybridization $V$.
The FSC phase, where
$\langle b_B \rangle\neq 0$, $\langle b_C\rangle \neq 0$, occurs in the low-temperature region, while the FLM phase, where  $\langle b_B\rangle=\langle b_C\rangle =0$, is the high temperature one. Between the FSC and FLM phases is the SSC phase, where the magnetic impurities are strongly coupled with conduction electrons only at either sites $B$ or $C$, and all other magnetic impurity sites are effectively depleted from the lattice.
This depletion is a strong correlation effect by the site selective condensation of the slave bosons due to the asymmetry between $B$ and $C$ sites. When $\varepsilon_B = \varepsilon_C$, the SSC disappears. In this case the $B$ and $C$ sites are equivalent, and their symmetry cannot be broken. In Fig. \ref{fig4} we also plot the LDOS in the FSC and SSC phases.
In the FSC phase, the soft-gap hybridization band  still exists near the Fermi level in the gap between the two subbands of conduction electrons, similar to the one in the previous depleted case. In addition to the soft-gap hybridization band, two narrow subbands appear at the gap edge of the conduction electron subbands. They are the hybridization bands between the flat band of conduction electrons and magnetic impurities. The flat-band hybridization bands can be interpreted as
a consequence of the molecular Kondo effect of magnetic impurity in the flat band \cite{Thuy}. However, in contrast to the single impurity case, where the soft-gap conduction electrons are quenched from the molecular Kondo singlet formation \cite{Thuy}, the soft-gap conduction electrons at sites $B$ and $C$ still form the hybridization band with the magnetic impurities. In the SSC phase, where the boson condensation occurs only at sites $B$ or $C$, the soft-gap hybridization band is smeared out, while the two flat-band hybridization bands still remain. These features of the LDOS distinguish the SSC from the FSC phase. Measuring the LDOS one would detect the soft-gap hybridization band, as well as, distinguish the SSC and the FSC phases.

\section{Uniform hybridization}
In this section we consider the case $V_A=V_B=V_C\equiv V$, i.e. the hybridizations are the same at every lattice site. However, the asymmetry of the magnetic impurities still can be maintained by varying $\varepsilon_\alpha$. The impurity asymmetry between the $B$ and $C$ sites would give two different critical lines. They together with the critical line for the boson condensation at $A$ sites could form a rich phase diagram.

\subsection{Case $|\varepsilon_A| \geq |\varepsilon_B|=|\varepsilon_C|  $}
\label{sub1}

\begin{figure}[t]
\centering
\begin{tabular}{cc}
\includegraphics[width=0.45\textwidth]{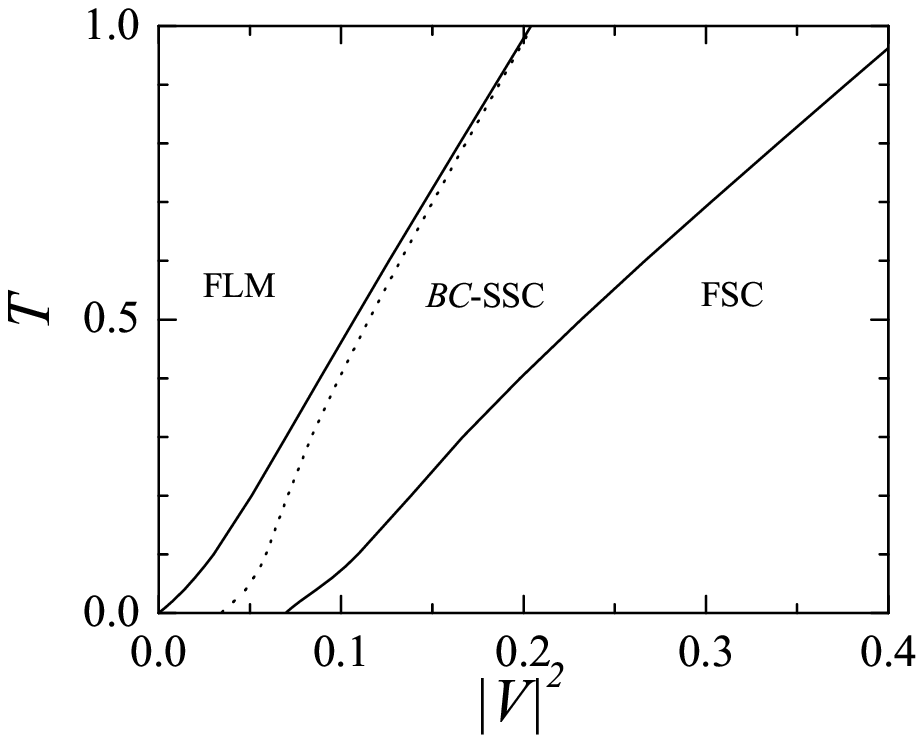} &
\includegraphics[width=0.45\textwidth]{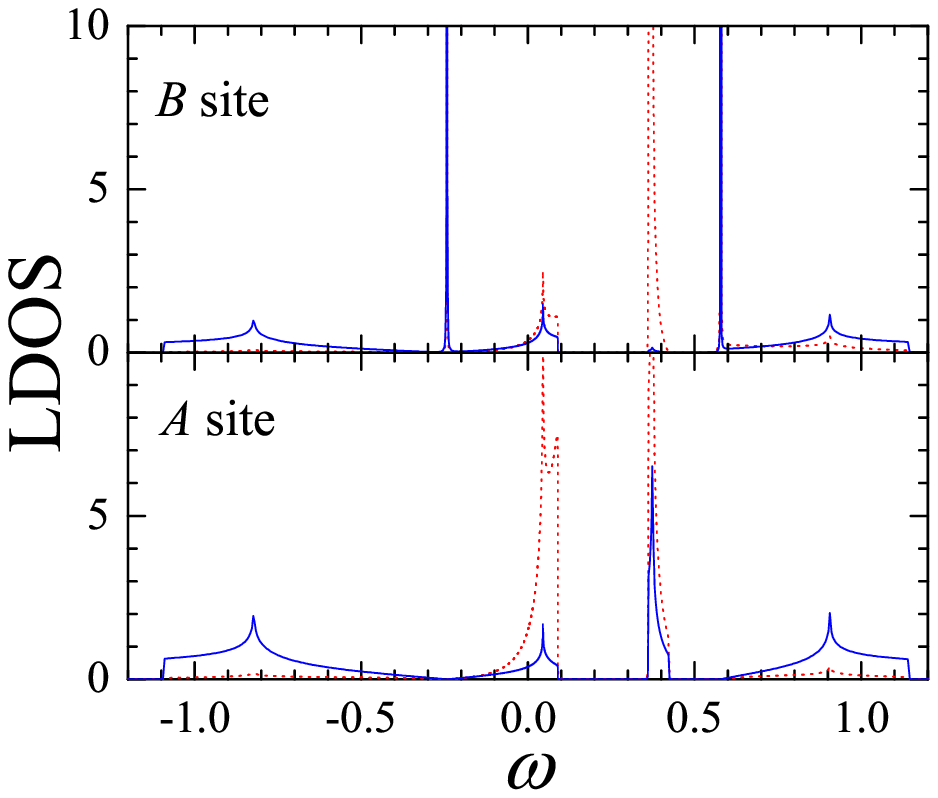}
\end{tabular}
\caption{Uniform hybridizations $V_A=V_B=V_C=V$. Left panel: Phase diagram for $\varepsilon_A=-0.2$, $\varepsilon_B=\varepsilon_C=-0.1$  (solid lines).  The dotted line is the phase boundary between the FSC and $BC$-SSC phases when $\varepsilon_A=\varepsilon_B=\varepsilon_C = -0.1$.
Right panel: LDOS of conduction electrons (solid line) and of magnetic impurities (dotted line) at $A$ and $B$ sites in the FSC phase for $V^2=0.25$,  $\varepsilon_A=-0.2$, $\varepsilon_B=\varepsilon_C=-0.1$ ($T=0.1$). The broadening parameter $\eta=10^{-4}$.
}
\label{fig6}
\end{figure}

In this case all edge-center sites are equivalent.
The phase diagram can be obtained from the critical lines determined by Eq. (\ref{cr}).
The critical lines separate the phase space $(V, T)$ into three regions as shown in Fig. \ref{fig6}.
At low temperature and strong hybridization,
$\langle b_B \rangle = \langle b_C \rangle \neq 0$ and $\langle b_A \rangle \neq 0$.
This yields the FSC state.
At high temperature and weak hybridization, $\langle b_A \rangle = \langle b_B \rangle = \langle b_C \rangle = 0$. This is the FLM phase. In the intermediate region, $\langle b_B \rangle = \langle b_C \rangle \neq 0$ and $\langle b_A \rangle = 0$. This is the $BC$-SSC phase. It is equivalent to the FSC of the depleted lattice with $V_A=0$. However, the depletion of the magnetic impurities at $A$ sites is a correlation effect, where the soft-gap Kondo effect at $A$ sites is absent. This depletion still occurs when $\varepsilon_A=\varepsilon_B = \varepsilon_C$, i.e. when all magnetic impurity parameters are uniform.
This selective Kondo SC at sites $B$ and $C$ is due to the flat-band singularity of the LDOS at sites $B$ and $C$.
The $BC$-SSC always exists for weak hybridizations regardless of the magnetic impurity parameters.
Actually, in the $BC$-SSC phase the molecular Kondo effect is dominant.
For $\varepsilon_A=\varepsilon_B = \varepsilon_C$, two phase boundaries, which separate the $BC$-SSC phase from the FSC and FLM ones, asymptotically approach each other at high temperature, as it is shown in Fig.~\ref{fig6}.

In Fig. \ref{fig6} we also plot the LDOS of conduction electrons and magnetic impurities at sites $A$ and $B$  in the FSC phase. The $BC$-SSC phase is equivalent to the FSC phase in the $A$-site depleted lattice, which we have already considered in Sec. \ref{secdepl}. In the FSC phase the LDOS at sites $A$ and $B$  exhibit different features. At $A$ sites the LDOS  exhibits two soft-gap hybridization bands, which locate in the gap between two conduction bands. Their appearance is due to the boson condensation at $A$ sites. Since the Kondo effect at $A$ sites is the soft-gap type, the LDOS at $A$ sites does not have the hybridization bands of the flat band and impurities. In contrast,
at $B$ or $C$ sites, two hybridization bands additionally occur at the gap edges due to the boson condensations at $B$ and $C$ sites. These hybridization bands have the same feature of the flat-band hybridization bands, which occur in the $A$-site depleted lattice.
These features are typical properties of the LDOS in the FSC phase.

\subsection{Case $|\varepsilon_A| < |\varepsilon_B|=|\varepsilon_C|  $}
\label{sub2}

\begin{figure}[t]
\centering
\begin{tabular}{cc}
\includegraphics[width=0.45\textwidth]{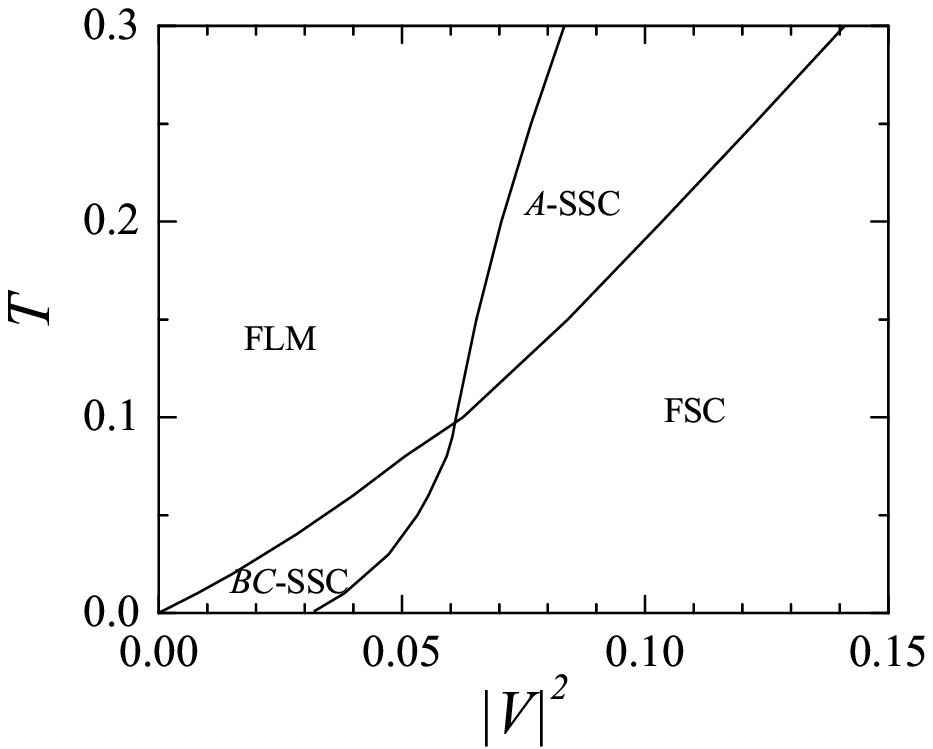} &
\includegraphics[width=0.45\textwidth]{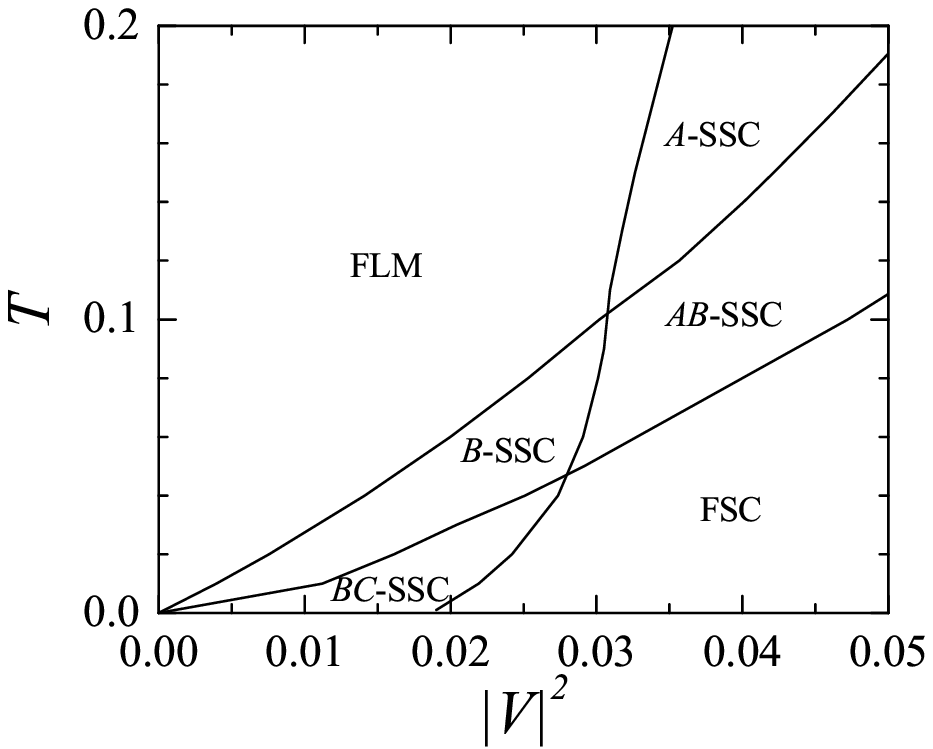}
\end{tabular}
\caption{Phase diagram for uniform hybridizations $V_A=V_B=V_C=V$. Left panel: case $\varepsilon_A=-0.1$, $\varepsilon_B=\varepsilon_C=-0.2$. Right panel: case $\varepsilon_A=-0.05$,  $\varepsilon_B=-0.1$,  $\varepsilon_C=-0.15$. }
\label{fig7}
\end{figure}

Like the previous subsection, in this case the impurity energy levels at $B$ and $C$ sites are the same, however, they lie lower than the impurities at $A$ sites. The phase diagram is plotted in the left panel of Fig. \ref{fig7}. In addition to the $BC$-SSC phase like the one in the previous subsection, the $A$-SSC phase appears in the high temperature and strong hybridization region. This $A$-SSC phase is equivalent to the one of the lattice with the depletion of magnetic impurities at $B$ and $C$ sites, since $\langle b_B \rangle = \langle b_C \rangle =0$. The $BC$-SSC phase exists only at low temperature and weak hybridization. The phase diagram also shows a quadruple critical point, where the FSC, FLM, $BC$-SSC and $A$-SSC phases coexist. However, this feature no longer exists when $\varepsilon_A = \varepsilon_B=\varepsilon_C$. This case (subsection \ref{sub2}) is not adiabatically connected to the previous one (subsection \ref{sub1}).

\subsection{Case $|\varepsilon_A| <|\varepsilon_B| < |\varepsilon_C|   $}
\label{sub3}

This case is also equivalent to the case $|\varepsilon_A| <|\varepsilon_C| < |\varepsilon_B|$. The phase diagram is very rich as it is shown in the right panel of Fig. \ref{fig7}. In addition to the FSC and FLM phases, both the two-site and single-site SSC phases appear. There are two quadruple critical points.
Unlike the previous cases (subsections \ref{sub1}-\ref{sub2}), the $BC$-SSC phase has the asymmetry $\langle b_B \rangle \neq \langle b_C \rangle$ due to $\varepsilon_B \neq \varepsilon_C$.
The $A$-SSC phase is equivalent to the one in the lattice with the impurity $B$ and $C$ sites depleted, whereas the  $B$-SSC phase
breaks the symmetry between $B$ and $C$ sites. In the $B$-SSC phase the boson condensation at $B$ sites is preferably selected due to
$|\varepsilon_B| < |\varepsilon_C|$. In the opposite case $|\varepsilon_C| < |\varepsilon_B|$, the boson condensation at $C$ sites
is selected instead.
A similar selection also occurs in the $AB$-SSC phase. However, when $\varepsilon_B = \varepsilon_C$, both the asymmetric $B$- and $AB$-SSC phases disappear.

\subsection{Case $|\varepsilon_B| <|\varepsilon_A| < |\varepsilon_C|   $}
\label{sub4}

\begin{figure}[t]
\centering
\begin{tabular}{cc}
\includegraphics[width=0.45\textwidth]{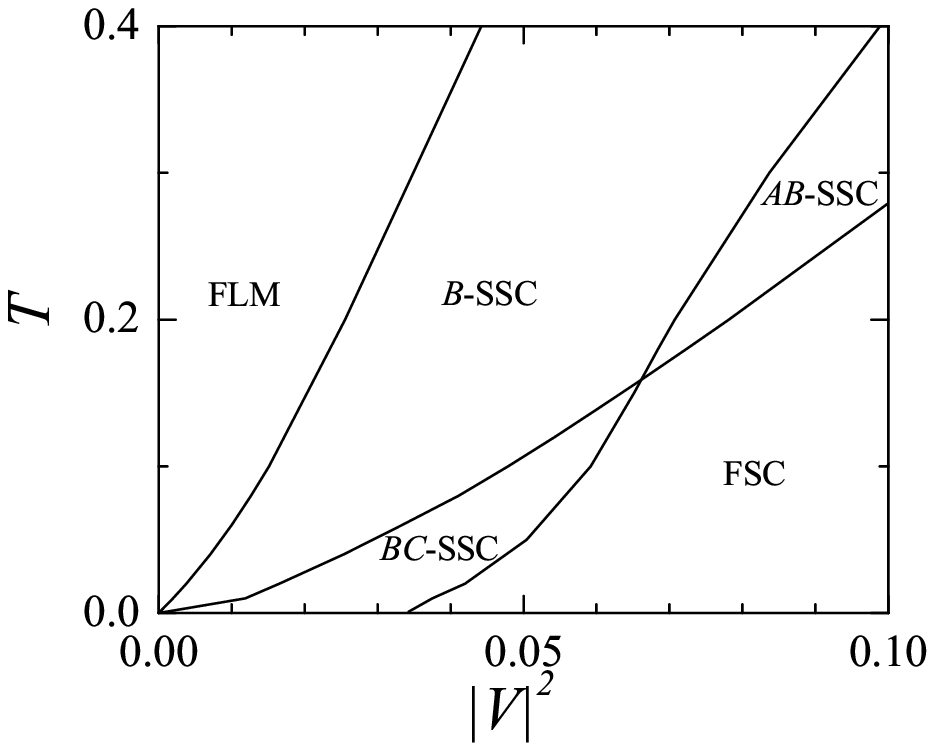} &
\includegraphics[width=0.45\textwidth]{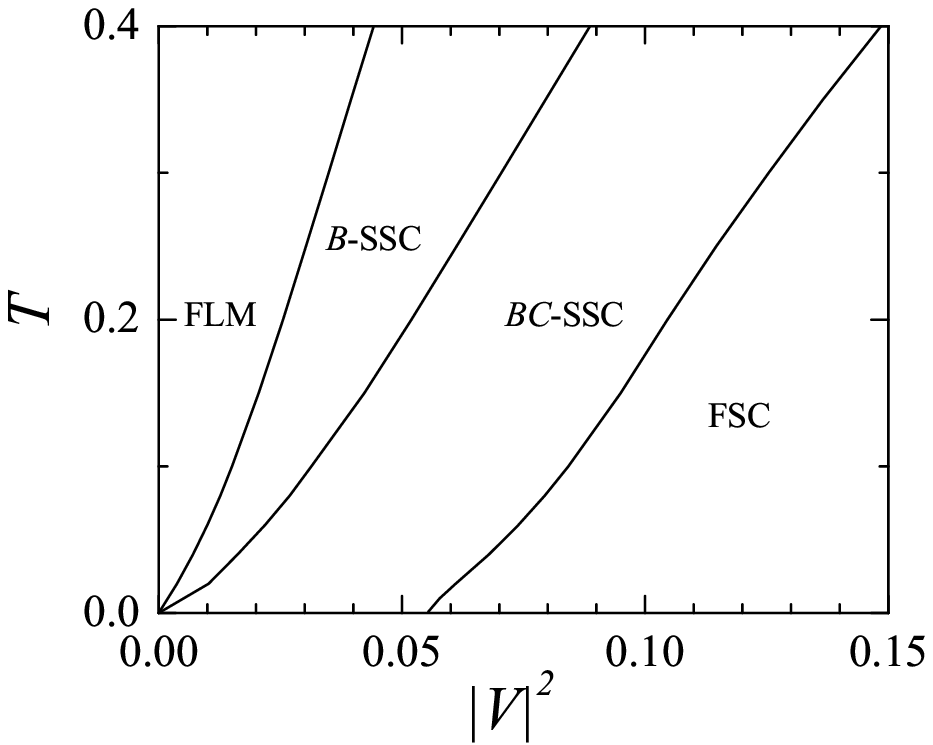}
\end{tabular}
\caption{Phase diagram for uniform hybridizations $V_A=V_B=V_C=V$. Left panel: case $\varepsilon_A=-0.1$, $\varepsilon_B=-0.05$,  $\varepsilon_C=-0.15$. Right panel: case $\varepsilon_A=-0.15$, $\varepsilon_B=-0.05$, $\varepsilon_C=-0.1$.}
\label{fig9}
\end{figure}

This case is also equivalent to the case $|\varepsilon_C| <|\varepsilon_A| < |\varepsilon_B|$. The phase diagram is plotted in the left panel of Fig. \ref{fig9}. There is a quadruple critical point. Similar to the previous case (subsection \ref{sub3}), the asymmetric $B$- or $AB$-SSC phases also occur due to $|\varepsilon_B| \neq |\varepsilon_C|$.
Since $|\varepsilon_B| < |\varepsilon_A| < |\varepsilon_C| $, the boson condensation at sites $B$ is more favorable than the one at sites $A$ and $C$. As a consequence, the Kondo SC at sites $B$ is always selected except in the FLM regime.
This feature is absent in the previous case, where $|\varepsilon_A| < |\varepsilon_B| $, in which the $A$-SSC phase occurs instead.
When  $|\varepsilon_C| <|\varepsilon_A| < |\varepsilon_B|$ the Kondo SC at sites $C$ is selected.

\subsection{Case $|\varepsilon_B| < |\varepsilon_C| \leq  |\varepsilon_A|$}
\label{sub5}

This case is also equivalent to the case $|\varepsilon_C| < |\varepsilon_B| \leq |\varepsilon_A|$.
The phase diagram is plotted in the right panel of Fig. \ref{fig9}.
Unlike the previous cases, there is no a quadrupole critical point. Since $|\varepsilon_B|$ and $|\varepsilon_C|$ are smallest among the energy levels of impurities, the Kondo SC at sites $B$ or $C$ are more favorable, and they are selected except for the FLM phase.
In Fig. \ref{fig9}, the $B$-SSC phase, which exists at high temperature and weak hybridization, is
selected due to $|\varepsilon_B| < |\varepsilon_C|$. There is a critical point, which separates the $BC$-SSC and FSC phases.

\section{Discussion and conclusion}
The PAM in the Lieb lattice exhibits very rich phase diagrams. Depending on the parameters of the magnetic impurities, different SSC phases occur.
However, at low temperature only the $BC$-SSC and FSC phases exist regardless of the impurity parameters. There is a critical point which separates the $BC$-SSC and the FSC phases.
Across their phase boundary, the magnetic impurities at $A$ sites are changed from the LM to the SC regimes, while the impurities at $B$ and $C$ sites  are always strongly coupled to conduction electrons. This selection of the Kondo SC at sites $B$ and $C$ indicates the domination of the flat-band Kondo effect at low temperature. Indeed, for a single impurity, the flat or narrow bands suppress the participation of other dispersive electrons in the Kondo-singlet formation, and as a consequence only the molecular Kondo effect occurs \cite{Thuy}. Only at enough strong hybridizations all magnetic impurities are strongly coupled to conduction electrons.
The selective Kondo SC is similar to the fractionalized Fermi liquid \cite{Senthil1,Senthil2}, because across the transition from the FSC to the $BC$-SSC phase, the Kondo screening at $A$ sites is broken down, leaving the local moments at $A$ sites decoupled.
One can also notice that the energy level of the magnetic impurities also plays a distinct role in the SSC. The magnetic impurities with deeper energy level (i.e. the absolute value of the energy level is larger) are less favorable to be selected for the Kondo SC regime. In particular, the magnetic impurities with deepest energy level are depleted from the lattice except for the FSC phase at strong hybridizations. With deeper energy level, the magnetic impurities are harder to be coupled with those conduction electrons, which are near the Fermi energy in order to form the Kondo singlet. However, in the SSC the band flatness plays a dominant role, because at low temperature the Kondo SC is selected at those lattice sites, where the LDOS has the flat-band feature, regardless of the energy level of magnetic impurities.

The selective Kondo SC is expected to occur in a number of systems.
First, it occurs in the magnetic impurity flat-band lattices. The perovskite lattice structure can be considered as a generation of the
Lieb lattice \cite{Tien,Weeks}. One can expect the selective Kondo SC occurs in heavy fermion perovskites. It is similar to the partial localization in heavy fermions of Uranium compounds, where both itinerant and localized $5f$ states coexist \cite{ZF,ZF1}. In the partial localization
electrons in orbitals with weak hybridization are selected to be localized \cite{ZF,ZF1}.
Flat bands also exist in geometrically frustrated lattices, such as the kagome, checkerboard lattices \cite{Derzhko}. Recently, experiments showed the Ce moments in heavy fermions on the kagome-like lattice CePdAl are partially screened by the Kondo physics \cite{Oyamada,Sakai}. This is closely related to the selective Kondo SC. Actually, the electron structure of the tight-binding model on
the kagome lattice also has a flat band and two Dirac-cone bands \cite{LiWang}. However, in contrast to the Lieb lattice, in the kagome lattice the flat band does not touch the crossing point of two linearly dispersing bands. A possibility of the selective Kondo SC in the kagome lattice requires a further study.
Second, the selective Kondo SC can occur in topological semimetals with triply band crossing when magnetic impurities are doped, because their low-energy effective Hamiltonian is identical to the one of the tight-binding model in the Lieb lattice \cite{Bradlyn}. Although, the band flatness in the triply band crossing semimetals exists only nearby the high-symmetric crossing point, it could still distinguish the orbital with band flatness from the other orbitals with the linear dispersion. As a consequence, the LDOS of these orbitals would be different, and this leads to the orbital SSC phases. The SSC may also occur in the 6- and 8-fold degenerate band semimetals too \cite{Bradlyn}. Although these
semimetals may not feature the band flatness, their orbital LDOS are still qualitatively different, and this may also lead to the orbital SSC phase when magnetic impurities are doped. Third, the selective Kondo SC  can also occur in lattices with nonuniform coordination number. In the Lieb lattice, the $A$ sites have the coordination number $z=4$, while at the $B$ or $C$ sites $z=2$. The nonuniform coordination number also occurs in various lattice structures, such as the body-centered-square lattice, the dice lattice, or in general the decorated lattices, depleted lattices, quasicrystals. Although the lattices with the nonuniform coordination number may not feature the band flatness, the nonuniform of the coordination numbers may lead the LDOS at different sites to be qualitatively different. As a result, the orbital SSC may be stabilized at low temperature.

In this paper we have found the selective Kondo SC in flat-band lattices.
However, we have only considered the non-magnetic phases and the interplay between the band flatness and the Kondo singlet formation.
The finite Coulomb interaction of the impurities
can result in a magnetic stability. Especially, when electron hopping between magnetic impurity sites is included, it together with the Coulomb interaction can generate the flat-band ferromagnetism \cite{Tasakirev}.
The interlay between the band flatness, magnetism and the Kondo effect would result into complexity which deserves research attention.

\section*{Acknowledgement}

This research is funded by Vietnam National Foundation
for Science and Technology Development (NAFOSTED) under Grant No 103.01-2017.13.

\end{document}